\documentstyle[amssymb,aps]{revtex}

\begin{document}
\title{Semiclassical Concepts in Magnetoelectronics}
\author{Gerrit E. W. Bauer, Yuli V. Nazarov, Daniel Huertas-Hernando,}
\address{Delft University of Technology, Department of Applied Physics and DIMES, \\
Lorentzweg 1, 2628 CJ Delft, The Netherlands}
\author{Arne Brataas}
\address{Harvard University, Lyman Laboratory of Physics, Cambridge, MA 02138}
\author{Ke Xia and Paul J. Kelly}
\address{Twente University, Department of Applied Physics 7500 AE Enschede, \ The\\
Netherlands}
\date{\today }
\maketitle

\begin{abstract}
Semiclassical theories of electron and spin transport in metallic magnetic
structures are reviewed with emphasis on the role of disorder and electronic
band structures in the current perpendicular to the interface plane (CPP)
transport configuration.
\end{abstract}

\pacs{Magnetoelectronics, theory, CPP, GMR, disorder, interface resistance}

\section{Introduction}

Electron transport in layered magnetic systems can be directed parallel or
perpendicular to the interfaces. The physics of transport in the latter,
so-called current perpendicular to the plane (CPP) configuration, studied
first by the Michigan State University Group with superconducting contacts 
\cite{Pratt1} and by Gijs {\it c.s.} in microstructured pillars at arbitary
temperatures \cite{Gijs}, has been reviewed quite recently \cite{Review}.
The topic remains to attract the interest of the community and new insights
have been obtained on issues as first-principles calculations of transport
in disordered multilayers, phase coherence effects, transport with
non-collinear magnetizations and spin-torques induced by applied currents.
The present manuscript briefly reviews and compares these novel
developments. Related topics of interest are superconductor-ferromagnet
hybrids \cite{Belzig00,Zareyan00}, spin-injection into carbon nanotubes \cite
{Tsukagoshi99,Balents00} and semiconductors \cite{Fiederling99} or
many-terminal devices \cite{Jedema00}.

Most studies have been carried out on multilayers consisting of many bilayer
periods. In earlier studies attention was focussed on collinear
magnetization profiles, {\it i.e.} that with antiparallel or parallel
magnetization vectors for neighbouring layers. These experiments are well
fitted by the so-called 2-channel series resistor model (2CSRM) which is
discussed in the light of recent developments in Chapter 2. In Chapter 3
novel theoretical approaches are reviewed which describe non-collinear
configurations.

\section{Collinear magnetization}

The experiments of perpendicular transport in magnetic multilayers are well
described by the 2CSRM, {\it i.e.} an equivalent electric circuit of two
spin channels in parallel, in which the resistance of each channel is the
sum of the bulk and interface resistances. When all magnetizations are
parallel the total resistance $R_{T}$ of a ferromagnetic/normal metal (F/N)
multilayer of $M$ double layers reads 
\begin{equation}
AR_{T}=M\left[ \rho ^{\left( N\right) }d_{N}+\sum_{s}\left( \rho
_{s}^{\left( F\right) }d_{F}+2AR_{s}^{N/F}\right) \right] ,  \label{srm}
\end{equation}
where A\ is the cross section of the sample, $\rho ^{\left( N\right) }$ is
the resistivity of the bulk normal metal,$\ \rho _{s}^{\left( F\right) }$ is
the resistivity of the ferromagnet for spin direction $s,$ $d_{N}$ and $%
d_{F} $ are the layer thicknesses and $R_{s}^{N/F}$ are the spin-dependent
interface resistances. The\ five parameters $\rho ^{\left( N\right) },$ $%
\rho _{s}^{\left( F\right) },$ and $R_{s}^{N/F}$ can be determined
accurately by fitting experiments on numerous samples with different layer
thicknesses and for antiparallel as well as parallel magnetic configurations 
\cite{Pratt1}. It was soon realized that the microscopic basis was to be
found in semiclassical arguments \cite{Levy91,Bauer92}. The most flexible
theoretical framework turned out to be the linearized Boltzmann equation in
the relaxation time approximation by Valet and Fert \cite{VF}, which also
included spin-flip processes. In this model the distribution function or
local chemical potential $f_{s}$ for spin $s$ in the $F$ or $N$ bulk
materials is governed by the one-dimensional diffusion equation: 
\begin{equation}
\frac{\partial ^{2}f_{s}(x)}{\partial x^{2}}=\frac{1}{2}\frac{%
f_{s}(x)-f_{-s}(x)}{l_{sf}^{2}}.\;  \label{valetfert2}
\end{equation}
where $\sigma _{s}$ is \ the conductivity and spin-flip processes are
included in terms of the spin-diffusion length $l_{sf}$. Interfaces were
treated as thin regions with a different (low) mobility. The current for
spin $s$ then reads: 
\begin{equation}
j_{s}(x)=\sigma _{s}\frac{\partial f_{s}(x)}{\partial x},
\end{equation}
When $l_{sf}\gg $ $d$ the results are identical to the 2CSRM. However, this
treatment is incomplete in that the discontinuities in the electronic
structure at heterointerfaces are disregarded, which are essential for the
electron transport properties since they scatter electrons even in clean
samples \cite{Schep95,Zahn95}. This can be seen most easily for an isolated
interface in a constriction according to the Landauer-B\"{u}ttiker formula : 
\begin{equation}
G^{A/B}=\frac{1}{R^{A/B}}=\frac{e^{2}}{h}\sum_{k_{\parallel }\nu
s,k_{\parallel }^{\prime }\nu ^{\prime }s^{\prime }}\left| t_{k_{\parallel
}\nu s,k_{\parallel }^{\prime }\nu ^{\prime }s^{\prime }}\right| ^{2}=\frac{%
e^{2}}{h}\sum_{\mu \mu ^{\prime }}T_{\mu \mu ^{\prime }}  \label{ir}
\end{equation}
where $t_{\mu \mu ^{\prime }},T_{\mu \mu ^{\prime }}$ are the transmission
coefficients and probabilities of states $\mu =k_{\parallel }\nu s$ at the
Fermi energy with (tranverse) wave vector $k_{\parallel }$ parallel to the
interface and band index $\nu $. For a ballistic interface $k_{\parallel }$
and $s$ are conserved during scattering. This expression can be compared
with that for a homogeneous point contact of the materials A and B with\
(Sharvin) conductances $G^{A}>G^{B}$. It is clear that any mismatch in the
electronic structure will reduce the conductance $G^{A/B}<G^{A}$.

\subsection{Interface resistance}

Of interest is the microscopic explanation of the interface resistance
parameter $R^{A/B}$ in Eq. (\ref{srm}) and its relation with Eq. (\ref{ir}).
The success of the 2CSRM provides the guidance for a quantitative
understanding of the experiments. The parameters which fit so many
experiments turn out to be universal for a given material combination. The
absence of a measurable dependence on the geometrical parameters $d_{N}$ and 
$d_{F}$ (when everything is kept the same) is a strong indication that
quantum interference terms are negligibly small. For the large thicknesses
which have predominantly been studied experimentally, this is not
surprising. But also for significantly exchange-coupled multilayers, quantum
well states should not significantly affect transport since disorder causes
a large semiclassical background. A semiclassical model therefore should be
appropriate for all but the cleanest samples with very thin layers. A simple
yet efficient {\it first principles }procedure to incorporate the interfaces
comes down to chopping the sample into slices of bulk layers and interfaces
which scatter electrons, separated by fictitious non-scattering regions \cite
{Schep97}. The bulk layer scattering can be treated in terms of transmission
and reflection matrices, modelled in two simple limits; the ballistic limit,
in which no scattering occurs during the transmission through the bulk and
the diffuse limit in which the scattering is isotropic, which means that the
transmitted electrons do not retain any memory of the incident wave vectors.
The transmission probabilities for the combined system then follows by the
semiclassical concatenation of transmission and reflection probability
matrices of the resistive elements in series{\it \ }\cite{Cahay88,Brataas94}%
. Indeed, the 2CSRM is recovered with interface resistances 
\begin{equation}
R^{A/B}=\frac{h}{e^{2}}\frac{1}{\sum T_{\mu \mu ^{\prime }}}-\frac{1}{2}%
\left( \frac{1}{G^{A}}+\frac{1}{G^{B}}\right) .  \label{Rdiff}
\end{equation}
where we have corrected for the spurious geometrical Sharvin resistances 
\cite{corr}. For ballistic bulk layer transmission \cite{Schep97} 
\begin{equation}
\left[ R^{A/B}\right] ^{-1}=\frac{e^{2}}{h}\sum_{\mu \nu }\left[ \left( {\bf %
I}-{\bf T}-{\bf R}\right) ^{-1}{\bf T}\right] _{\mu \nu },  \label{Rball}
\end{equation}
which is in fact the ``old'' Landauer formula. Both Eqs. (\ref{Rdiff}) and (%
\ref{Rball}) do not contain any parameters and provide a first-principles
prescription for the semiclassical interface resistances. They hold not only
in the metallic case but should be also valid when the interface is in the
tunneling regime. Stiles and Penn \cite{Stiles00} confirmed that the
expression (\ref{Rdiff}) agrees very well with the solutions of the
Boltzmann equation.

Levy {\it et al.} \cite{Levy00} recently found strong effects of disorder in
the leads\ on transport through tunnel junctions. Our discussion implies
that these effects should be absent in a semiclassical description, which is
at odds with the conclusions\ of Levy {\it et al.}. Interference effects
between impurities and resonance states at the tunnel junction might provide
an explanation \cite{Ke}.

\subsection{First-principles calculations}

The interface resistance has been calculated by first principles for
specular interfaces in Refs. \cite{Schep97,Stiles00,Xia00} for ballistic
magnetic domain walls in Ref. \cite{Hoof99} and diffuse interfaces in Ref. 
\cite{Xia00}. In the table we summarize our results for specular and rough
interfaces, the latter modelled as a 50\%/50\% bilayer alloy, and compare
them with available experiments. Remarkable is the different behaviour of
the Co%
\mbox{$\vert$}%
Cu as compared with the Fe%
\mbox{$\vert$}%
Cr interface with respect to the interface roughness: the spin contrast of
the Co%
\mbox{$\vert$}%
Cu interface transparency is weakly enhanced by disorder, but strongly
reduced in Fe%
\mbox{$\vert$}%
Cr. The message for experimentalists is that in Fe%
\mbox{$\vert$}%
Cr it should pay off to optimize the epitaxial growth parameters.

The Boltzmann equation has been solved numerically by Butler {\it et al.} 
\cite{Butler00} for Co%
\mbox{$\vert$}%
Cu%
\mbox{$\vert$}%
Co perpendicular spin valves (see also \cite{Chien99}). The constant
relaxation time approximation for the bulk materials was used and
distribution functions were matched via the transmission and reflection
coefficients for specular interfaces. They found results for the interface
resistance slightly different from that of \cite{Schep97,Xia00}. Butler {\it %
et al.} find a dependence of the interface resistance on small Cu
thicknesses, which appears to contradict the universality of the parameters
of the 2CSRM. These corrections are not the small quantum size effects found
by Xia {\it et al.} \cite{Xia00} and by Tsymbal {\it et al.} \cite
{Bozec00,Tsymbal00}, but are purely classical effects due to evanescent
terms in the distribution functions. These can be interpreted as the
corrections to the assumption of complete isotropy by Schep {\it et al. } 
\cite{Schep97}. The isotropy conditions is expected to be much better
fulfilled when interface disorder is taken into account. It is clear that
more, also experimental, work is needed to understand transport for very
thin layer thicknesses in which the incomplete randomization and the
appearance of quantum corrections will cause deviations from the two-channel
resistor model.

\section{Non-collinear magnetization}

Electron transport in devices with non-collinear magnetizations have come
into focus by the recent interest in the torque exerted on the magnetization
by a spin-polarized injected current \cite{Katine00}. The theories for
collinear magnetization mentioned above have been extended to the case in
which the magnetizations are not collinear,{\it \ i.e.} not parallel or
antiparallel, namely via a magnetoelectronic circuit theory \cite{Brataas00a}%
, and for CPP spin-valve structures, random matrix theory of transport \cite
{Waintal} as well as a direct solution of the diffusion equation in the
presence of an external magnetic field \cite{Huertas00}. Recently also
exchange-biased CPP spin-valves are under scrutiny \cite{Pratt00}. We find
that these approaches are equivalent and reduce to previous theories in the
limit of collinear magnetization. An interesting point is that \cite
{Brataas00a,Waintal} do not start from the outset with a semiclassical
approximation, but derive it from an isotropy assumption, thus put previous
more or less {\it ad hoc} approaches on firm theoretical foundations.

\subsection{Circuit theory}

Transport in hybrid metallic systems can be described by a generalization of
Kirchhoff's theory of electronic circuits when parts of the system are not
phase-coherently coupled. This approach has been pioneered in Ref. \cite
{Nazarov} for electronic networks with superconducting elements. It has
recently been adopted also for magnetoelectronic circuits \cite{Brataas00a},
like the Johnson spin transistor \cite{Johnson} or the 4-terminal mesoscopic
spin valve of Jedema {\it et al.} \cite{Jedema00}. For a different approach
to the many terminal magnetolectronic circuits see \cite{Geux00,Tang00}. The
circuit theory can be derived from a given Stoner Hamiltonian in terms of
the Keldysh non-equilibrium Green function formalism in spin space \cite
{Brataas00b}. The basic physics is provided by splitting up the system into
reservoirs, resistors and nodes, where the latter can be real or fictitious
(as discussed above). In order to arrive at a useful formalism, an isotropy
assumption has to be introduced for the nodes, namely that the electron
distributions in the nodes are isotropic, which implies sufficient disorder
(or chaotic scattering) to allow for configurational averaging. It does not
require any inelastic\ or dephasing scattering mechanism, although, when
happening in the nodes, they will not hurt either. Because the
spin-accumulation is not necessarily parallel to the spin-quantization axis,
at each node the electron distribution at a given energy $\epsilon $ can be
denoted as $\hat{f}(\epsilon ),$ where the hat ($\hat{}$) denotes a $2\times
2$ matrix in spin-space. The external reservoirs are assumed to be in local
equilibrium so that the distribution matrix is diagonal in spin-space and
attains its local equilibrium value $\hat{f}=\hat{1}f(\epsilon ,\mu _{\alpha
})$, $\hat{1}$ is the unit matrix, $f(\epsilon ,\mu _{\alpha })$ is the
Fermi-Dirac distribution function and $\mu _{\alpha }$ is the local chemical
potential in reservoir $\alpha $. The direction of the magnetization of the
ferromagnetic nodes is denoted by the unit vector ${\bf m}_{\alpha }$. The
current through each contact can be calculated as a function of the
distribution matrices on the adjacent nodes $2\times 2$ conductance tensors.
The current matrix (for an F%
\mbox{$\vert$}%
N junctions) reads 
\begin{equation}
e\hat{I}=G^{\uparrow }\hat{u}^{\uparrow }\left( \hat{f}^{F}-\hat{f}%
^{N}\right) \hat{u}^{\uparrow }+G^{\downarrow }\hat{u}^{\downarrow }\left( 
\hat{f}^{F}-\hat{f}^{N}\right) \hat{u}^{\downarrow }-G^{\uparrow \downarrow }%
\hat{u}^{\uparrow }\hat{f}^{N}\hat{u}^{\downarrow }-(G^{\uparrow \downarrow
})^{\ast }\hat{u}^{\downarrow }\hat{f}^{N}\hat{u}^{\uparrow }\,,
\label{curgen}
\end{equation}
in terms spin-%
$\frac12$%
rotation matrices $\hat{u}^{s}=\left( \hat{1}+s{\bf \hat{\sigma}}\cdot \vec{m%
}\right) /2$, distribution matrices $\hat{f}^{F}$,$\hat{f}^{N}$ on
ferromagnetic and normal node, the spin-dependent conductances $G^{\uparrow
} $ and $G^{\downarrow }$ (which in planar junctions should be corrected as
discussed in Section II) 
\[
G^{s}=\frac{e^{2}}{h}\left[ M-\sum_{nm}|r_{s}^{nm}|^{2}\right] =\frac{e^{2}}{%
h}\sum_{nm}|t_{s}^{nm}|^{2}\,, 
\]
and the mixing conductance 
\begin{equation}
G^{s,-s}=\frac{e^{2}}{h}\left[ M-\sum_{nm}r_{s}^{nm}(r_{-s}^{nm})^{\ast }%
\right] ,  \label{Gmixing}
\end{equation}
where $r_{s}^{nm},$ $t_{s}^{nm}$ are the reflection and transmission
coefficients, $M$ the number of modes in the absence of reflections.
Spin-flips in the contacts have been disregarded. The spin-current
conservation law 
\begin{equation}
\sum_{\alpha }\hat{I}_{\alpha \beta }=\left( \frac{\partial \hat{f}_{\beta }%
}{\partial t}\right) _{\mbox{rel}}\,=\frac{\hat{1}\mbox{Tr}\hat{f}^{N}-2\hat{%
f}^{N}}{2\tau _{sf}}  \label{curcons}
\end{equation}
allows computation of the circuit properties as a function of the applied
voltages, where $\hat{I}_{\alpha \beta }$ denotes the $2\times 2$ current in
spin-space from node (or reservoir) $\alpha $ to node (or reservoir) $\beta $
and the term on the right hand side describes spin-relaxation in the
(normal) node. The right hand side of Eq.~(\ref{curcons}) can be set to zero
when the spin-current in the node is conserved, {\em i.e.} when an electron
resides on the node sufficiently shorter than the spin-flip relaxation time $%
\tau _{sf}$.

Some insight can be gained by re-writing the current and the distribution
function in the form of a scalar particle and a vectorial spin contribution, 
$\hat{I}=(I_{0}+\text{\boldmath$\sigma $}\cdot {\bf I}_{s})/2$, $\hat{f}%
^{N}=f_{p}^{N}+\text{\boldmath$\sigma $}\cdot {\bf s}\Delta f^{N}$ and $\hat{%
f}^{F}=f_{p}^{F}+\text{\boldmath$\sigma $}\cdot {\bf m}\Delta f^{F}$. The
spin-current through an F%
\mbox{$\vert$}%
N interface can then be expanded into different vector components as: 
\begin{eqnarray}
{\bf I}_{s} &=&{\bf m}[(G^{\uparrow }-G^{\downarrow
})(f_{p}^{F}-f_{p}^{N})+(G^{\uparrow }+G^{\downarrow })\Delta
f^{F}+(G^{\uparrow }+G^{\downarrow }-2\text{Re}G^{\uparrow \downarrow }){\bf %
s}\cdot {\bf m}\Delta f^{N}]  \nonumber \\
&&{\bf s}2\text{Re}G^{\uparrow \downarrow }\Delta f^{N}+({\bf s}\times {\bf m%
})2\text{Im}G^{\uparrow \downarrow }\Delta f^{N}\,.  \label{spincurrent}
\end{eqnarray}
The vector spin current component perpendicular to the magnetization
direction equals the spin-torque exerted by the polarized current on the
ferromagnet \cite{Katine00,Waintal}.

\subsection{Random matrix theory}

Waintal {\it et al.} \cite{Waintal} have extended the random matrix theory
of transport \cite{Beenakker1} to include non-collinear magnetizations in F%
\mbox{$\vert$}%
N%
\mbox{$\vert$}%
F junctions. This paper is focussed on the spin torque, but the physics is
essentially the same as for transport in a disordered CPP spin valve with
arbitrary magnetization configuration as discussed above. The objects which
are averaged are the scattering matrices of the bulk layer of the normal
metal assuming that all members of the ensemble fulfil the symmetry
requirements of the problem and are equally probable. The theory is more
intricate than in the case of superconducting S%
\mbox{$\vert$}%
N%
\mbox{$\vert$}%
S junctions \cite{Beenakker1}, because the averaging has to be carried out
over the eigenvalues and eigenvectors. Analytical results can be obtained
for the leading term of an expansion into $1/N$, {\it i.e.} the inverse of
the number of transport channels.

It is easily seen that the analytical relations obtained by Waintal {\it et
al.} \cite{Waintal} for halfmetallic ferromagnetic elements agree with those
from the circuit theory for the symmetric two-terminal device. The general
case is less obvious, but by somewhat tedious manipulations a complete
equivalence of the final equations for both theories can be proven \cite
{Arne00}. We may conclude from Waintal's results that what seemed to be an
assumption in the procedure of Schep {\it et al. }\cite{Schep97}, {\it viz}.
the semiclassical concatenation, can be {\it derived} from the isotropy
assumption. Although not yet worked out, Waintal {\it et al.'s} approach can
be generalized to include quantum corrections, which become important for a
small number of channels, as well as many terminal configurations. It is not
clear how spin-flip-relaxation processes can be incorporated, which is quite
straight-forward for the circuit theory.

\subsection{Diffusion equation for non-collinear transport}

In some cases the theories above are not sufficient and the spatially
dependent ditribution has to be evaluated. The spin-polarized electron
distribution is characterized by a $2\times 2$ matrix in spin space of the
form: 
\begin{equation}
\hat{f}^{N}(x)=\left( 
\begin{array}{cc}
\begin{array}{c}
f_{\uparrow \uparrow }^{N}(x)
\end{array}
& 
\begin{array}{c}
f_{\uparrow \downarrow }^{N}(x)
\end{array}
\\ 
\begin{array}{c}
\\ 
f_{\downarrow \uparrow }^{N}(x)
\end{array}
& 
\begin{array}{c}
\\ 
f_{\downarrow \downarrow }^{N}(x)
\end{array}
\end{array}
\right) .  \label{distribfN.}
\end{equation}
When the size of the system $L$ is larger than the spin diffusion length $%
l_{sf}$, $\hat{f}^{N}(x)$ depends on the position. We have studied transport
through an F%
\mbox{$\vert$}%
N%
\mbox{$\vert$}%
F device under the condition $l_{f}\ll l_{sf}$, where $l_{f}=v_{F}\left(
1/\tau +1/\tau _{sf}\right) ^{-1}$ is the{\em \ mean free path}, $v_{F}$ is
the Fermi velocity, $\tau $ the spin-conserving scattering time and $\tau
_{sf}$ the spin-flip scattering time \cite{Huertas00}. Under the condition $%
l_{f}\ll l_{sf}=\sqrt{v_{F}l_{f}\tau _{sf}/3}$, we obtain the generalized
diffusion equation in the normal metal 
\begin{equation}
\frac{\partial ^{2}\hat{f}^{N}(x)}{\partial x^{2}}=\frac{1}{l_{sf}^{2}}%
\left( \hat{f}^{N}(x)-{\bf \hat{1}}\frac{\mbox{Tr}\left( \hat{f}%
^{N}(x)\right) }{2}\right) -\frac{i}{\hslash }\left[ \frac{g\mu _{B}}{2D}%
\left( {\bf \hat{\sigma}\cdot }\vec{B}\right) ,\text{ }\hat{f}^{N}(x)\right]
_{-}.
\end{equation}
Its solution, with boundary conditions at the interface governed by the
conductance matrix, describes {\it e.g.} the precession of the
spin-accumulation in an applied magnetic field and lead to a physical
interpretation of the imaginary part of the mixing conductance \cite
{Huertas00}.

\subsection{CPP spin valve}

The different approaches described above lead to an analytical expression
for the total conductance of CPP spin valves as a function of the angle
between the magnetizations of the different ferromagnets $\theta $, when $%
l_{sf}\gg L$, at zero magnetic field ($\vec{B}=0$) and for symmetric
contacts: 
\[
{G^{T}(\theta )}=\frac{G}{2}\left( 1-P^{2}\frac{\tan ^{2}\theta /2}{\tan
^{2}\theta /2+\frac{\left| \eta \right| ^{2}}{%
\mathop{\rm Re}%
\eta }}\right) 
\]
where $G=G^{\uparrow }+G^{\downarrow },$ $P=\left( G^{\uparrow
}+G^{\downarrow }\right) /G,$ 
\begin{equation}
\eta =\frac{2G^{\uparrow \downarrow }}{G};\;\frac{\left| \eta \right| ^{2}}{%
\mathop{\rm Re}%
\eta }=\frac{4\left| G^{\uparrow \downarrow }\right| ^{2}}{G^{2}}\frac{G}{2%
\mathop{\rm Re}%
G^{\uparrow \downarrow }}=\frac{2\left| G^{\uparrow \downarrow }\right| ^{2}%
}{G%
\mathop{\rm Re}%
G^{\uparrow \downarrow }}
\end{equation}
The angular magnetoconductance reads: 
\begin{equation}
\frac{{G^{T}(\theta )-G^{T}(0)}}{{G^{T}(\pi )-G^{T}(0)}}=\frac{\tan
^{2}\theta /2}{\tan ^{2}\theta /2+\frac{\left| \eta \right| ^{2}}{%
\mathop{\rm Re}%
\eta }}=\left\{ 
\begin{array}{c}
\sin ^{2}\theta /2 \\ 
\frac{%
\mathop{\rm Re}%
\eta }{\left| \eta \right| ^{2}}\tan ^{2}\theta /2
\end{array}
\right. \text{ for } 
\begin{array}{c}
\eta =%
\mathop{\rm Re}%
\eta \rightarrow 1 \\ 
\eta \rightarrow \infty
\end{array}
\end{equation}
Balents and Egger \cite{Balents00} arrive at the same result in their study
of spin-injection into carbon nanotubes in the non-interacting limit. The
electron-electron interaction is found to enhance the mixing conductance.

\section{Discussion}

Our understanding of the transport properties of the CPP multilayers is
semi-quantitative for the parallel aligned and the as grown ``virgin''
samples, in which to a good approximation neighbouring magnetization vectors
are antiparallel. The basis of this understanding is (1) knowledge of the
magnetization configuration and (2) the 2CSRM. Bozec {\it et al.} \cite
{Bozec00} claimed to have found evidence for a breakdown of the 2CSRM.
However, a recent study with intentionally alloyed bulk layers comes to
different conclusions, {\it i.e.} that the 2CSRM should remain unchallenged
for collinear magnetic structures, be they ``type I'' or ``type II'',
interleaved or separated \cite{Eid00}.

The situation of the magnetic-field cycled ``deflowered'' samples is more
difficult. The experiments of Bozec {\it et al.} \cite{Bozec00} could be
explained by a ``spin-memory'' effect caused by spin-flip at the interfaces
which can be incorporated into the two-channel series resistor model \cite
{Eid00}. This picture requires that the magnetization at all intermediate
fields is random but essentially collinear (except possibly at interfaces).
An alternative explanation is Wiser's hypothesis that the angle between the
magnetizations of different layers is rotated during magnetization reversal 
\cite{Bozec00}. The transport properties based on this hypothesis can be
computed in principle by the generalizations of the 2CSRM to non-collinear
transport discussed above, which we may call ``matrix series resistor
model''. At the moment the magnetization distribution is not known
sufficiently well, but it seems likely that non-collinearity and randomness
both play a role. Exchange-biased CPP spin valves appear to be better suited
to test the new theories than multilayers \cite{Pratt00}.

\section{Acknowledgment}

We acknowledge discussions with Kees Schep, Wolfgang Belzig, Bill Pratt,
Jack Bass, Ilya Turek, Josef Kudrnovsk\'{y}, and Maciej Zwierzycki as well
as support by FOM and the NEDO joint research program (NTDP-98). A.B. is
supported by the Norwegian Research Council.


\begin{references}
\bibitem{corr}  This correction is necessary due to the fictitious nature of
the non-scattering regions and should be omitted in the case of resistive
elements like point contacts which separate real non-scattering islands.
\end{references}
\end{document}